\documentclass[fleqn,10pt]{wlscirep}
\usepackage{pifont}% http://ctan.org/pkg/pifont

\title{Impurity bound states in fully gapped $d$-wave superconductors with subdominant order parameters}

\author[1,*]{Mahdi Mashkoori}
\author[1]{Kristofer Bj\"{o}rnson}
\author[1]{Annica M. Black-Schaffer}
\affil[1]{Department of Physics and Astronomy, Uppsala University, Box 516, SE-751 20 Uppsala, Sweden}

\affil[*]{Correspondence and requests for materials should be addressed to M.M. (email: mahdi.mashkoori@physics.uu.se)}

\begin{abstract}
Impurities in superconductors and their induced bound states are important both for engineering novel states such as Majorana zero-energy modes and for probing bulk properties of the superconducting state. The high-temperature cuprates offer a clear advantage in a much larger superconducting order parameter, but the nodal energy spectrum of a pure $d$-wave superconductor only allows virtual bound states. Fully gapped $d$-wave superconducting states have however been proposed in several cuprate systems thanks to subdominant order parameters producing $d+is$- or $d+id'$-wave superconducting states. Here we study both magnetic and potential impurities in these fully gapped $d$-wave superconductors.
Using analytical T-matrix and complementary numerical tight-binding lattice calculations, we show that magnetic and potential impurities behave fundamentally different in $d+is$- and $d+id'$-wave superconductors. In a $d+is$-wave superconductor, there are no bound states for potential impurities, while a magnetic impurity produces one pair of bound states, with a zero-energy level crossing at a finite scattering strength. On the other hand, a $d+id'$-wave symmetry always give rise to two pairs of bound states and only produce a reachable zero-energy level crossing if the normal state has a strong particle-hole asymmetry. 
\end{abstract}

\begin{document}
\newcommand{\be}{\begin{equation}}
\newcommand{\ee}{\end{equation}}
\newcommand{\bearr}{\begin{eqnarray}}
\newcommand{\eearr}{\end{eqnarray}}
\newcommand{\nn}{\nonumber}
\newcommand{\la}{\langle}
\newcommand{\ra}{\rangle}
\newcommand{\cd}{c^\dagger}
\newcommand{\vd}{v^\dagger}
\newcommand{\fd}{f^\dagger}
\newcommand{\tk}{{\tilde{k}}}
\newcommand{\tp}{{\tilde{p}}}
\newcommand{\tq}{{\tilde{q}}}
\newcommand{\eps}{\varepsilon}
\newcommand{\vk}{{\vec k}}
\newcommand{\vp}{{\vec p}}
\newcommand{\vq}{{\vec q}}
\newcommand{\vkp}{\vec {k'}}
\newcommand{\vpp}{\vec {p'}}
\newcommand{\vqp}{\vec {q'}}
\newcommand{\bk}{{\bf k}}
\newcommand{\bp}{{\bf p}}
\newcommand{\bq}{{\bf q}}
\newcommand{\up}{\uparrow}
\newcommand{\down}{\downarrow}
\newcommand{\fns}{\footnotesize}
\newcommand{\cdag}{c^{\dagger}}
\newcommand{\lc}{\langle\!\langle}
\newcommand{\rc}{\rangle\!\rangle}

\flushbottom
\maketitle

\thispagestyle{empty}

\section*{Introduction}
It is well-established that a single magnetic impurity induces so-called Yu-Shiba-Rusinov (YSR) bound states inside the energy gap of conventional $s$-wave superconductors.\cite{Yu1965, Shiba1968, Rusinov1969}
In systems with spin-orbit coupling, such intra-gap bound states have recently been proposed to give rise to emergent Majorana zero-energy modes at the end-points of chains of magnetic impurities.\cite{Mourik2012,Yazdani2014,Franke2015PRL} This has lead to a surge of interest in impurity-induced bound states in superconductors, with both experimental and theoretical work focusing on properties ranging from high angular momentum scattering and complex internal structure of the impurities to quantum phase transitions and spontaneous current generation, as well as many other aspects.\cite{Sasha2006,Wang2012, Franke2015,Ruby2016,Sau2013,KristoferPRL,KristoferPRB2016, Kim2015,Kristofer2016PiShift}
In addition, the physical properties of an impurity give valuable information about the bulk itself and can thus be a decisive probe for establishing the properties of the bulk superconducting state.~\cite{slager2015}

One severely limiting factor in all these studies is the low superconducting transition temperature accompanied small energy gap associated with conventional $s$-wave superconductors. The cuprate superconductors with their much high transition temperatures would here be a tantalizing option, if it were not for their $d$-wave order parameter symmetry which enforces a nodal energy spectrum.\cite{Tsuei} 
The low-energy nodal quasiparticles prevent impurity bound states and thus a pure $d$-wave superconductor can only host virtual bound states.\cite{Sasha1995,Sasha2006} 
However, in small islands of YBa$_2$Cu$_3$O$_{7-\delta}$ a fully gapped spectrum has recently been discovered and attributed to the existence of subdominant order parameters, with the superconducting symmetry likely being either $d_{x^2-y^2}+is$-wave (d+is) or chiral $d_{x^2-y^2}+id_{xy}$-wave ($d+id'$).\cite{Floriana2013,Annica2013} Both of these subdominant orders produce a hard gap and spontaneously break time-reversal symmetry, since the free energy very generally is minimized for subdominant parameters with an overall $\pi/2$-phase shift relative to the dominant order. The $d+id'$-wave state is also a chiral state with its non-trivial topology classified  by a Chern number $N = 2$.\cite{Volovik97}
Evidence also exists that surfaces, especially the (11) surface,
\cite{Elhalel} 
as well as certain impurities
\cite{BalatskyPRL98} 
also spontaneously generate a time-reversal symmetry breaking superconducting state with either $d+is$- or $d+id'$-wave symmetry.

In this work we establish the properties of both potential and magnetic impurities in these two fully gapped $d$-wave superconductors. More specifically, we investigate the intra-gap bound states due to potential and magnetic impurities using both an analytic continuum T-matrix formulation and numerical tight-binding lattice calculations.
We show that impurities create entirely different 
bound states in $d+is$-wave and chiral $d+id'$-wave superconductors, despite both being fully gapped and with a dominant parent $d$-wave state.
These distinct behaviours of single impurities have direct consequences as to the possibility of utilizing $d$-wave superconductors for producing Majorana zero-energy modes in magnetic wires or islands constructed out of collections of single impurities. Moreover, in spite of being only localized imperfections to the lattice structure, we find that impurities are very suitable for probing and differentiate the symmetry of the superconducting state. This is in sharp contrast to the virtual bound states in pure $d$-wave states, which persist even above the superconducting transition temperature and consequently, can not be considered to be a good probe of the symmetry of the superconducting state.\cite{Hudson2008}

For a $d+is$-wave superconductor we find that a potential impurity does not induce any bound states, while a magnetic impurity gives rise to a pair of bound states, which behaves very similar to the YSR bound states in conventional $s$-wave superconductors. This includes the 
behaviour of the energy spectrum when tuning the scattering strength $U_{mag}$ of the magnetic impurity. Since the $d+is$-wave state is topologically trivial and with a low-energy $s$-wave gap, this resemblance with a conventional $s$-wave superconductor is very plausible. 
More specifically, we find that the magnetic impurity bound states have a zero-energy level crossing at a finite critical scattering $U^c_{mag}$.
We are able to extract an analytical expression for the critical coupling which depends only on the ratio of the dominant $d$-wave  to the subdominant $s$-wave order parameter.
Moreover, through self-consistent tight-binding calculations we find a first-order quantum phase transition at $U_{mag}^c$, which also induces a local $\pi$-phase shift for $s$-wave component of the order parameter.

For the chiral $d+id'$-wave state we find a very different behaviour. Here both potential and magnetic impurities induce two pairs of bound states. For superconductors with a particle-hole symmetric normal state, the bound states are two-fold degenerate and there is no level crossings for any finite coupling. Instead, it is only in the unitary scattering limit ($U^c\rightarrow \infty$) that the bound states approach the middle of the gap. 
Doping the normal state away from particle-hole symmetry, the degeneracy is lifted for a magnetic impurity but not for a potential impurity. Moreover, for finite doping there is now a zero-energy level crossing, but for low doping it occurs only at very large scattering strengths.
Self-consistent calculations for a single impurity in a $d+id'$-wave superconductor finds a first-order phase transition at the level crossing, but no local phase shifts either the dominant and subdominant order parameters.
Considering that recent experiments have demonstrated access to adjustable magnetic scattering strengths $U_{mag}$,\cite{Franke2015} magnetic 
impurities offer a very intriguing way to clearly distinguish between the chiral $d+id'$-wave state and the likewise time-reversal symmetry breaking but topologically trivial $d+is$-wave state.

\section*{Results}

\subsection*{Analytic T-matrix calculations}
Impurity-induced bound states only exist in $d$-wave superconductors with a fully gapped energy spectrum. Introducing a subdominant superconducting order parameter will achieve this, since it very generally align with a complex $\pi/2$ phase relative to the dominant $d$-wave state. Here we consider a two-dimensional (2D) $d_{x^2-y^2}$-wave superconducting state with a complex subdominant order parameter such that the order parameter takes the form of
 $\Delta(k) = \Delta_1(k)+i\Delta_2(k)$ $(\Delta_1,\Delta_2 \in \mathbb{R})$. 
 More specifically, we treat the two most likely candidates:  $d_{x^2-y^2}+is$- and $d_{x^2-y^2}+id_{xy}$-wave symmetries. In order to achieve a good analytical understanding of the effect of impurities we here first perform T-matrix calculations. Later, we confirm and extend these results by also performing self-consistent tight-binding lattice calculations.

The Hamiltonian in presence of a single impurity (magnetic and/or potential)  reads (using $\hbar = 1$)
 \begin{equation}
\begin{array}{c}
{H_{BdG}} = \sum\limits_k {\psi _k^\dag
[\xi(k){\tau _3}{\sigma_0} - \Delta_1(k) {\tau _2}{\sigma_2}-\Delta_2(k)\tau_1\sigma_2]
{\psi _k}}, \\
{H_{imp}} = \sum\limits_{kk'\sigma }{\psi _k^\dag \hat{V} {\psi _{k'}}} 
= \sum\limits_{kk'\sigma }{\psi _k^\dag [U_{pot}\tau_3\sigma_0+U_{mag}\tau_3\sigma_3] {\psi _{k'}}},
\end{array}
 \end{equation}
where we use the Nambu  space spinor
$\psi _k^T =({\begin{array}{*{10}{c}}{{c_{k \uparrow }}}&{{c_{k \downarrow }}}&{c_{ - k \uparrow }^\dag }&{c_{ - k \downarrow }^\dag}\end{array}})$. 
Here $U_{pot}$ and $U_{mag}$ are the potential and magnetic scattering matrix elements induced by the impurity, while the kinetic energy is $\xi(k)$. The exact form of $\xi(k)$ is unimportant as we can linearise the spectrum around the Fermi level, setting $\xi(k) \approx v_{F} (k - k_F)$. \cite{MyNote1} 
The dominant order parameter is $\Delta_1(k)$, while $\Delta_2(k)$ represents the subdominant order parameter.
We assume that the magnetic moment is large enough to ignore quantum fluctuations and thus
we treat the impurity as a classical spin. The local moment of the impurity is directed along the $z$ easy axis, but it is straightforward
to show that the results are not affected by this assumption, since the electrons pair in the spin-singlet channel.
The matrices $\tau_i$ and $\sigma_i$ are the Pauli matrices acting  in particle-hole and spin spaces, respectively, while
 $\tau_0$ and $\sigma_0$ are unit matrices.
The bare Green's function for the superconductor is
\be\hat G_k^0\left( \omega  \right) = \frac{\omega \hat 1 + \xi(k) \tau_3\sigma_0 -\Delta_1(k) {\tau _2}{\sigma_2}-\Delta_2(k)\tau_1\sigma_2 }{\omega^2 - E^2(k)}, \ee
with the energy spectrum $E(k) =\pm  \sqrt{\xi^2(k) + |\Delta(k)|^2 }$. 
The Green's function in presence of a single impurity then reads
$ \hat G(k,k',\omega)  = {\delta _{kk'}}\hat G_k^0\left( \omega  \right) + \hat G_k^0\left( \omega  \right)\hat{T}\left( \omega  \right)\hat G_{k'}^0\left( \omega  \right)$, 
with the T-matrix $
\hat T\left( \omega  \right) = [{{1 - \hat V\sum\limits_k^{} {\hat  G_k^0\left( \omega  \right)} }}]^{-1}\hat V$.\cite{Mahan2013} Therefore,
finding the roots of the denominator of the T-matrix gives the energy of the impurity-induced bound states. For these analytical calculations we assume that the order parameter does not notably depend on the \textit{magnitude} of 
wave vector $k$, but only on its direction $\Delta(\phi) = \Delta_1(\phi) + i\Delta_2(\phi)$, but note that this assumption is not needed in the numerical lattice calculations. \cite{MyNote1}

Let us first consider a fully particle-hole symmetric spectrum for the normal state, which imposes $\sum \xi(k) = 0$. To access the T-matrix denominator the summation over the bare Green's function is needed,
$\sum\limits_k^{} {\hat G_k^0\left( \omega  \right)}  = 
 {F_0}\left( \omega  \right) \hat 1 +F_1(\omega) \tau_2 \sigma_2 + F_2(\omega)  {\tau _1}{\sigma_2}$, where we have defined $F_i(\omega)$ as
\be
\begin{array}{l}
{F_0}\left( \omega  \right) =   \frac{\rho }{2}\int_0^{2\pi } {d\phi \frac{-\omega}{{\sqrt {|\Delta(\phi)|^2 - {\omega ^2}} }}} ; \ \
{F_1}\left( \omega  \right) =  \frac{\rho }{2}\int_0^{2\pi } {d\phi \frac{\Delta_1(\phi)}{{\sqrt {|\Delta(\phi)|^2 - {\omega ^2}} }}} ; \ \
{F_2}\left( \omega  \right) = \frac{\rho }{2}\int_0^{2\pi } {d\phi \frac{\Delta_2(\phi)}{{\sqrt {|\Delta(\phi)|^2 - {\omega ^2}} }}}.
\end{array}
\label{disSG0.eq}
\ee
Here $\rho = {k_F}/({2\pi v_F})$ is the density of states of the 2D free electron gas at the Fermi level.
The above result is for a particle-hole symmetric normal state, but this symmetry is often broken in reality. We use the chemical potential $\mu$ to measure the degree of particle-hole symmetry breaking in the energy spectrum, thus leaving the case $\mu = 0$ to represent full particle-hole symmetry.
Considering a small deviation from particle-hole symmetry, such that $\mu/\Lambda \ll 1$ where $\Lambda$ is the energy integration cut-off,
 the summation of the bare Green's function also contains the term $F_3\tau_3\sigma_0$. Up to first order in $\mu/\Lambda$ we find $F_3 = 2\rho{\mu}/{\Lambda}$,~\cite{MyNote2} while it is straightforward to show that in this limit $F_0$, $F_1$, and $F_2$ remain unchanged from the particle-hole symmetric case.

\subsubsection*{$d+is$-wave state}
\begin{figure}[hbt]
\center
\includegraphics[scale=0.65]{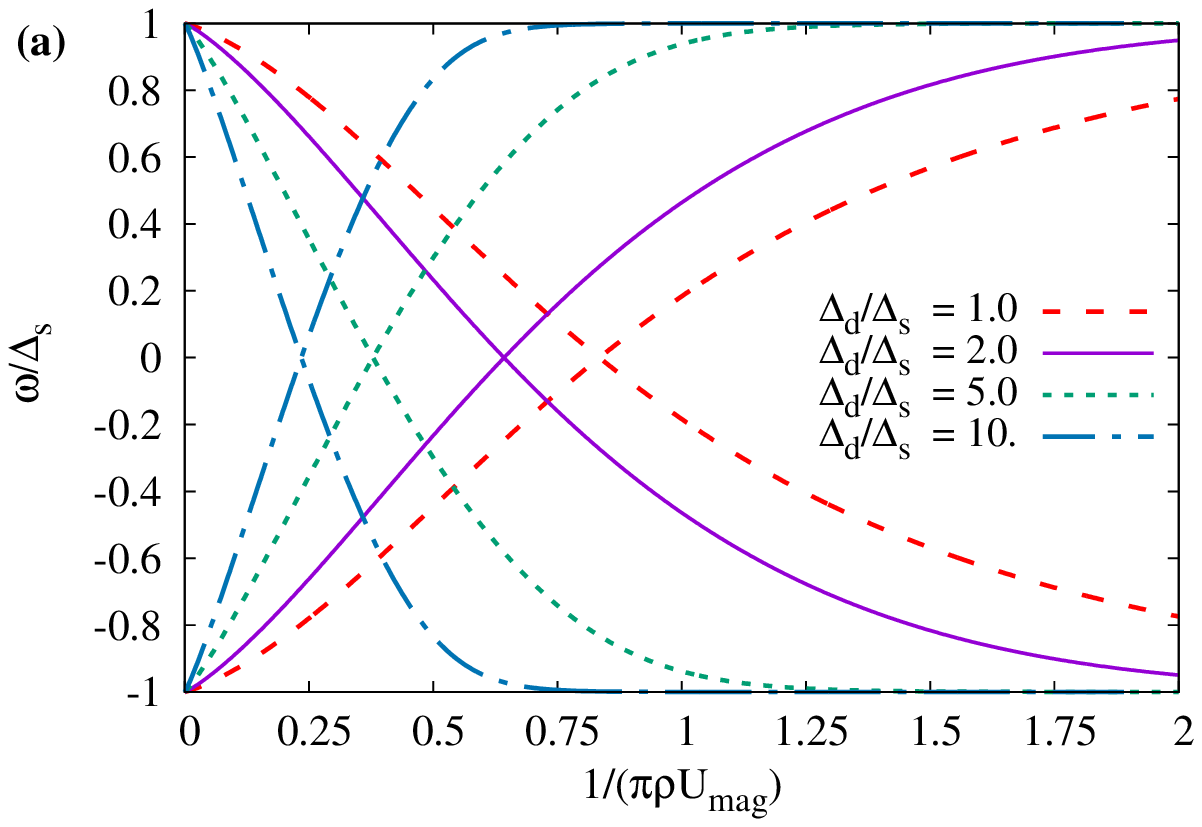}
\includegraphics[scale=0.65]{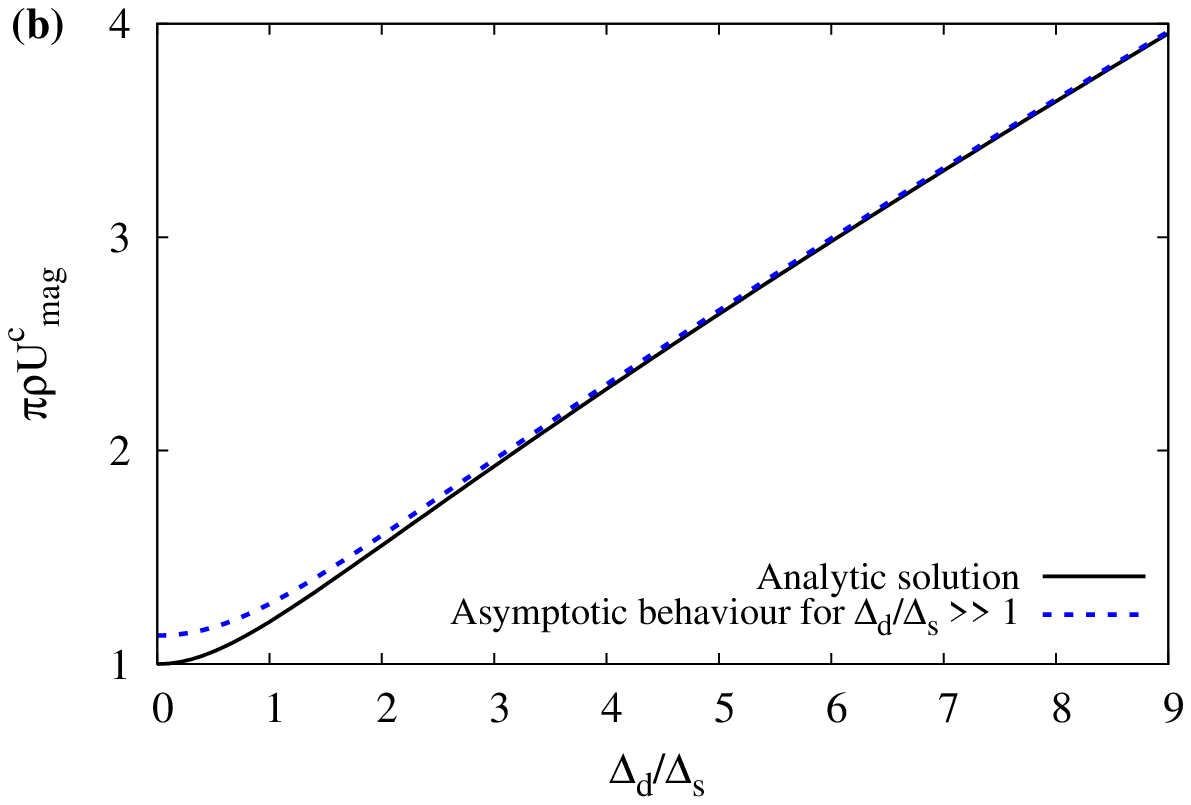}
\caption{The energy spectrum of all intra-gap bound states for a magnetic impurity in a $d+is$-wave supercondutor as a function of $\tilde{U}_{mag}^{-1}$ (a) and the critical magnetic scattering $\tilde{U}_{mag}^c$ for the zero-energy level crossing as function of $\tilde{\Delta}=\Delta_d/\Delta_s$ (b).}
\label{d+is.fig}
\end{figure}

First, we consider the $d_{x^2-y^2}+is$ state where the order parameter is of the form $\Delta(\phi) = \Delta_{d}\cos(2\phi) + i\Delta_{s}$. In this case $F_1(\omega) = 0$, due to the periodicity of the cosine function and the subdominant order parameter $\Delta_s$ not depending on $\phi$.  
Then, in the limit of $\mu = 0$, the bound states can be found as the solutions to
\be
\left\{ {1 \pm 2{U_{mag}}{F_0}\left( \omega  \right) - \left( {U_{pot}^2 - U_{mag}^2} \right)\left[ {F_0^2\left( \omega  \right) - F_2^2\left( \omega  \right)} \right]} \right\} = 0.
\ee
Since we are interested in real bound states, we only look for solutions $\omega \in \mathbb{R}$.
Further, to make sure that the bound states are isolated from the continuum spectrum of the superconducting quasiparticles, we limit ourselves to solutions that lie inside the gap, i.e.~$\omega \in \left[-\Delta_2,\Delta_2\right]$.
For a purely potential impurity we find no bound states, while for a purely
magnetic impurity there is one pair of solutions that do not depend on the 
sign of $U_{mag}$.
In order to find the bound state energies for a magnetic impurity we rephrase  $F_0(\omega)$ and $F_2(\omega)$ in terms of the complete elliptic integral of the first kind $K$,\cite{Abramowitz1964handbook} resulting in
\be
\frac{1}{\tilde{U}_{mag}} = \frac{{2\left( {1 \pm \tilde{\omega} } \right)}}{{\pi \sqrt {\tilde{\Delta} ^2 + 1 - {\tilde{\omega} ^2}} }}K\left( {\frac{\tilde{\Delta} ^2}{{\tilde{\Delta} ^2 +1 - {\tilde{\omega} ^2}}}} \right),
\label{BS-dis.eq}
\ee
where we have defined $\tilde{U}_{mag} =  \pi \rho U_{mag}$, 
$\tilde{\Delta} = {\Delta_d}/{\Delta_s}$, and $\tilde{\omega} = {\omega}/{\Delta_s}$.
For $\tilde{\Delta} = 0$, we naturally recover the YSR bound states found in a conventional $s$-wave superconductor: 
$\tilde{ \omega} = \pm  ({1-\tilde{ U}_{mag}^2})/({1 + \tilde{U}_{mag}^2})$.~\cite{Yu1965,Shiba1968,Rusinov1969}
The bound state spectrum for a general $d+is$-wave superconductor comes as the solution of Eq.~\eqref{BS-dis.eq} and is illustrated in Figure~\ref{d+is.fig}(a).
As the figure shows, there is one pair of intra-gap bound states appearing at the gap edges for a weak magnetic impurity and moving toward the middle of the gap, such that at a critical magnetic scattering $U^c_{mag}$ a level crossing occurs. This behaviour is qualitatively
similar to the YSR bound pair in a conventional $s$-wave superconductor, where the level crossing signals a quantum phase transition between two different ground states.
For a magnetic impurity stronger than $U^c_{mag}$, the ground state will have one unpaired
electron because it is energetically favoured by the system to break a Cooper-pair to
 partially screen the impurity.~\cite{SalkolaPRB1997,Franke2015} The same quantum phase transition takes place also in the $d+is$-wave superconductor. Interestingly, for a $d+is$-wave superconductor, the critical coupling depends only on $\tilde{\Delta}$ and not on $\Delta_d$ and $\Delta_s$ separately.
By setting $\tilde{\omega} = 0$ in Eq.~\eqref{BS-dis.eq}, we find the analytically exact expression 
 $\tilde{U}^c_{mag}
 = {\pi\sqrt{1+\tilde{\Delta}^2}}/[2K \left({\tilde{\Delta}^2}/({1+\tilde{\Delta}^2})\right)]$.
Assuming $\tilde{\Delta} \gg 1$, the critical coupling reads $\tilde{U}^c_{mag} \approx \pi \sqrt{1+\tilde{\Delta}^2}/ \ln[16(1+\tilde{\Delta}^2)]$, which, as seen in Figure~\ref{d+is.fig}(b), only deviates at small $\tilde{\Delta}$ from the exact result.
This clearly illustrates that, in order to find zero modes, a larger moment and/or coupling is needed when the subdominant $s$-wave parameter is small compared to the $d$-wave order.

If we now break the electron-hole symmetry of the normal state, i.e.~assume $\mu \neq 0$, 
the bound states are instead found as the solution of 
\be
\begin{array}{l}
\left\{ {1 \pm 2{U_{mag}}{F_0}\left( \omega  \right) - 2{U_{pot}}{F_3} + \left( {U_{pot}^2 - U_{mag}^2} \right)\left[ {F_2^2\left( \omega  \right) + F_3^2-  F_0^2\left( \omega  \right)} \right]} \right\} = 0.
\end{array}
\ee
It is clear that also in this case, for a purely potential impurity 
 there are no real roots and consequently, potential impurities never induces any 
 bound states in a $d+is$-wave superconductor.
Moreover, the modifications of the bound state spectrum induced by a magnetic impurity is of the order $(\mu/\Lambda)^2$
and the change to the critical coupling is also of the same order of magnitude and thus negligible.~\cite{Mynote3}

\subsubsection*{$d+id'$-wave state}
\begin{figure}[h]
\center
\includegraphics[scale=0.65]{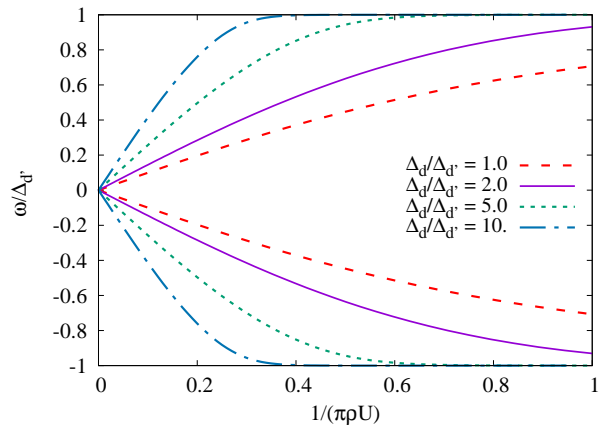}
\caption{The energy spectrum of all intra-gap bound states for a purely magnetic or potential 
impurity in a chiral $d$-wave superconductor as function of $\tilde{U}^{-1}$, representing either a magnetic or potential scattering matrix element. }
\label{fig2.fig}
\end{figure}

Next, we turn to the bound state formation inside the gap of a chiral $d$-wave or $d_{x^2-y^2} + id_{xy}$-wave superconductor. Here the order parameter is $\Delta(k) = \Delta_d\cos(2\phi)+\Delta_{d'}\sin(2\phi)$ and because of the periodicity of $\cos(2\phi)$ and $\sin(2\phi)$, all  off-diagonal terms in the summation of the bare Green's function vanish. Left for a particle-hole symmetric normal state spectrum $(\mu =0)$ is then only
\be
{F_0}\left( \omega  \right) \equiv  \frac{\rho }{2}\int_0^{2\pi } {d\phi \frac{-\omega }{{\sqrt {\Delta _d^2{{\cos }^2}\left( {2\phi } \right) + \Delta _d'^2\sin^2({2\phi}) - {\omega ^2}} }}} =  - \frac{{2\rho \omega }}{{\sqrt {\Delta _d^2 - {\omega ^2}} }}K(\frac{{\Delta _d^2 - \Delta _d'^2}}{{\Delta _d^2 - {\omega ^2}}}).
\label{didSG0.eq}
\ee
In this case the bound states are found as the solutions to
\be
\left[ {1 \pm 2{U_{pot}}{F_0}\left( \omega  \right) + \left( {U_{pot}^2 - U_{mag}^2} \right)F_0^2\left( \omega  \right)} \right] = 0.
\ee
Very interestingly, pure potential or pure magnetic impurities in a chiral $d$-wave superconductor lead to exactly the same pairs of intra-gap bound states as 
shown in Figure~\ref{fig2.fig}. 
In earlier work it has been claimed that the number of bound states for a potential impurity is only two.\cite{Wang2012} However, according to our results, these bound states are doubly degenerate, and there are in total four bound states. This statement is valid for both potential and magnetic impurities. Staying at $\mu = 0$ we find that for a potential impurity the negative energy branch (those occupied at zero temperature) consist of one spin up and one spin down state, and there is thus a Kramers degeneracy of the states.
However, for a magnetic impurity the bound states with negative energy are both spin down quasiparticles.
For an impurity with both potential and magnetic scattering
effects on the charge carriers, four non-degenerate bound states are generally present in a chiral $d$-wave superconductor and the two-fold degeneracy present in Fig.~\ref{fig2.fig} is lifted.

If the particle-hole symmetry of the normal states is broken, here by setting $\mu \neq 0$, then the summation over the bare Green's function also contains $F_3\tau_3\sigma_0$,
where, up to first order in $\mu/\Lambda$, $F_3 = 2\rho {\mu}/{\Lambda}$. 
The influence of this new term on the bound states is much more pronounced for the chiral $d$-wave state compared to the $d+is$-wave state.
\begin{figure}[htb]
\center
\includegraphics[scale=0.6]{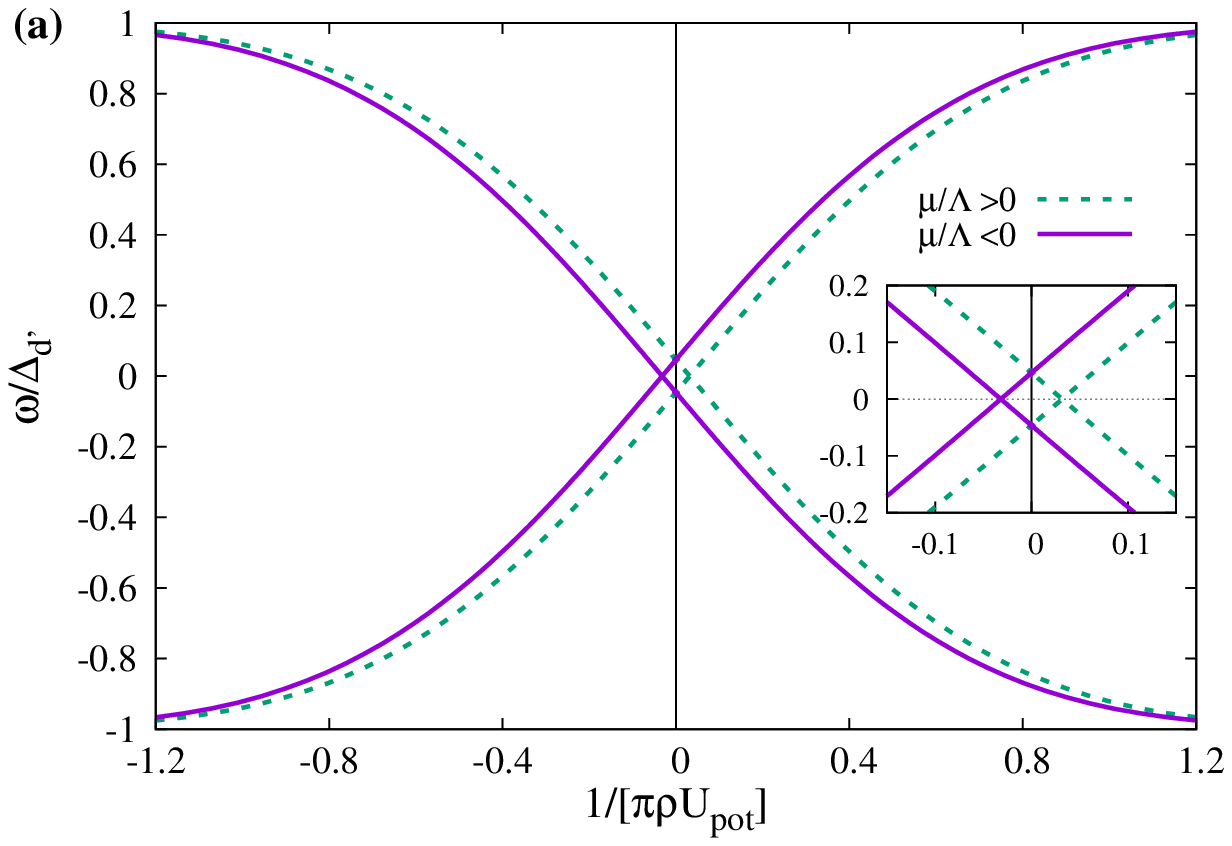}
\includegraphics[scale=0.6]{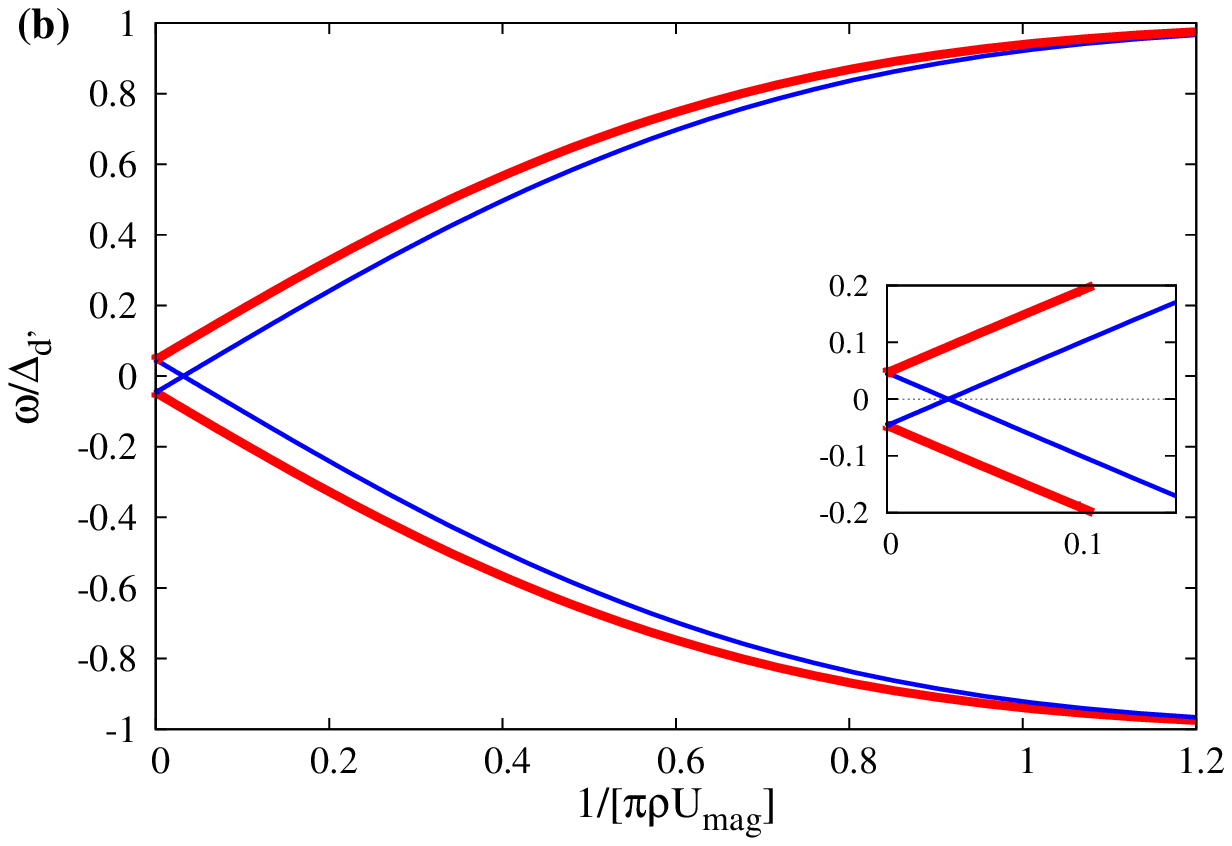}
\caption{The energy spectrum of all intra-gap bound states for a potential (a) and magnetic (b) impurity in a chiral $d$-wave superconductor at finite doping away from half-filling ($|\mu| = 0.1$).
Here $\Delta_d/\Delta_{d'} = 2$ and with the same vertical and horizontal axes as in Figure~\ref{fig2.fig}. }
\label{did-NPH.fig}
\end{figure}
For a potential impurity the bound states energies are still two-fold degenerate, 
but a non-zero $\mu$ 
shifts the energy of the bound states in the unitary limit, as is illustrated in Figure~\ref{did-NPH.fig}(a).
In fact, for the electron doped case, $\mu>0$, a level crossing appears for repulsive impurity scattering $(U_{pot} >0)$, while for a hole doped system, $\mu < 0$, this instead occurs for an attractive impurity $(U_{pot} <0)$.
Thus doping can be used as a simple means to control the level crossing for potential impurities. A local chemical potential induced by a tunneling probe could even offer in-situ tunability of the level crossing.\cite{Sau2013}

For a magnetic impurity even the bound state degeneracy is lifted for finite $\mu$, as seen in Figure~\ref{did-NPH.fig}(b).
In this case, one pair of bound states move away from the middle of the gap (thick red lines). For these states no zero modes are expected even in the unitary scattering limit for positive scattering $U_{mag} >0$. However, for negative scattering a zero energy bound states will appear because the bound states are always 
symmetric under $U_{mag} \rightarrow -U_{mag}$. For this reason we only plot the bound states for positive scattering in Figure~\ref{did-NPH.fig}(b). 
Regarding the other pair (thin blue lines), it
moves toward the middle of the gap and consequently a level crossing appears at $U_{mag}^c > 0$. In addition to being symmetric with respect to the sign of $U_{mag}$, the bound states also do not depend on the sign of $\mu$. 
Remarkably, the dimensionless critical coupling for reaching a zero-energy state for both magnetic and potential scatterers is the same $|\tilde{U}^c | = \pi \Lambda/|\mu|$.
As seen, this critical coupling can be decreased by increasing $|\mu|$.

For the purpose of the forthcoming self-consistent numerical calculation, we mention already here that even at half-filling, the energy degeneracy can still be lifted by adding a small amount of extended $s$-wave superconductivity
to the chiral $d$-wave. More precisely, if the order
parameter takes the form $\Delta(k) = \Delta_d\cos(2\phi) + i\Delta_{d'}\sin(2\phi) + \Delta_{s'}$, where $\Delta_{s'}$
is $k$-independent, the energy degeneracy for magnetic impurity bound states is lifted,
while the bound states remain degenerate for a potential impurity. 
Therefore, the degeneracy of the bound states in the presence of a magnetic impurity in a chiral $d$-wave superconductor is very fragile and can very easily 
be lifted.

\subsection*{Numeric tight-binding lattice calculation}
We now turn to discuss the results obtained from tight-binding lattice calculations.
In all calculations we use a generic finite-size square lattice in which we consider a single impurity located at the middle site. Again we consider both a $d+is$- and chiral $d+id'$-wave superconductor. The effective Hamiltonian for the 2D superconducting host with an impurity with both potential and magnetic scattering elements located at $R$ is~\cite{Annica2013,Sato2010}
\be
\begin{array}{l}
H_{BdG} =  - t\sum\limits_{\left\langle {{\bf{i}},{\bf{j}}} \right\rangle \sigma }^{} {c_{{\bf{i}}\sigma }^\dag {c_{{\bf{j}}\sigma }} + } \sum\limits_{\left\langle {{\bf{i}},{\bf{j}}} \right\rangle }^{}\frac{1}{4} {{\Delta _d}\left( {{\bf{i}},{\bf{j}}} \right)\left[ {c_{{\bf{i}} \uparrow }^\dag c_{{\bf{j}} \downarrow }^\dag  - c_{{\bf{i}} \downarrow }^\dag c_{{\bf{j}} \uparrow }^\dag } \right]} 
 + \sum\limits_{\bf{i}}^{} {{\Delta _s}\left( {\bf{i}} \right)c_{{\bf{i}} \uparrow }^\dag c_{{\bf{i}} \downarrow }^\dag }  + \sum\limits_{\left\langle {\left\langle {{\bf{i}},{\bf{j}}} \right\rangle } \right\rangle }^{} 
\frac{1}{4}{{\Delta _{d'}}\left( {{\bf{i}},{\bf{j}}} \right)\left[ {c_{{\bf{i}} \uparrow }^\dag c_{{\bf{j}} \downarrow }^\dag  - c_{{\bf{i}} \downarrow }^\dag c_{{\bf{j}} \uparrow }^\dag } \right]} + {\rm H.c.}, \\
 H_{imp} = \sum\limits_{\sigma \sigma'} \left[ U_{pot} c_{R \sigma}^{\dag} (\sigma_0)_{\sigma \sigma'} c_{R \sigma'} + U_{mag}   c_{R \sigma}^{\dag} (\sigma_z)_{\sigma \sigma'} c_{R \sigma'}  \right].
 \label{TBH}
\end{array}
\ee 
Here $\textbf{i} = (i_x,i_y)$ represents a site in the square lattice, with the lattice spacing $ a $ set to be 1. The dominant $d$-wave order exists on nearest neighbour bonds, while the subdominant $s$-wave order is an on-site parameter and the $d'$-wave order reigns on next nearest neighbour bonds. For the self-consistent calculations (see below) we do not a priori assume any  symmetries or conditions for any of these three order parameters. However, in calculations with constant order parameters throughout the sample, i.e.~non-self-consistent calculations, we enforce the $d$-wave order by setting  $\Delta_d((i_x,i_y),(i_x\pm1,i_y)) = -\Delta_d((i_x,i_y),(i_x,i_y\pm1))$ for all sites, i.e.~the order parameter on $y$-directed bonds are equal in magnitude but with opposite sign compared to the order parameter on $x$-directed bonds. Likewise, the $d'$-wave state has opposite signs on bonds in the $\pm(x+y)$ direction compared to in the $\pm(x-y)$ direction. We also by hand enforce the subdominant order parameter ($s$ or $d'$) to be purely imaginary in the non-self-consistent calculations.

We solve Eq.~\eqref{TBH} by performing a Chebyshev polynomial expansion of the corresponding Green's function.~\cite{ChebyshevRevModPhys,Covaci,Kristofer2016}
This method allows us to investigate lattices with very large number of  lattice points because the amount of necessary computational resources grow only linearly with the size of the system, far outperforming regular diagonalization. More specifically, we calculate the Green's function for the impurity site and its closest neighbours. The imaginary part of Green's function gives the local density of states (LDOS) and the bound states are easily identified as sharp peaks inside the energy gap at the impurity site and also its neighbouring sites.

\begin{figure}[htb]
\center
\includegraphics[scale=0.65]{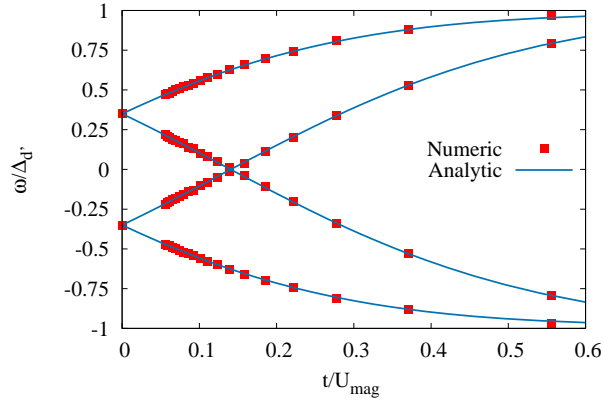}
\caption{The energy spectrum of all intra-gap bound states for a magnetic impurity in a chiral $d$-wave superconductor as function of inverse of impurity strength $U^{-1}_{mag}$. The order parameters are set to $\Delta_{d'}/t  = 0.05$,  $\Delta_d/\Delta_{d'}=2$, and $\mu/t=0.5$, and with lattice size $201 \times 201$.}
\label{Chebyshev-did-compare.fig}
\end{figure}

We find that the tight-binding calculations fit exactly to the analytical T-matrix results. For instance, in Figure \ref{Chebyshev-did-compare.fig} we show the bound states spectrum generated by a magnetic impurity in a chiral $d$-wave superconductor at finite doping, for both the numeric tight-binding method and an analytic T-matrix calculation. 
 In order to be able to do this comparison, we evaluate the summations appearing in the T-matrix formalism for a discrete mesh
over the first Brillouin zone of the square lattice using the same form of the kinetic energy.\cite{Mynote4}
We also find an excellent agreement in the 
unitary scattering limit $(U_{mag} \rightarrow \infty)$, which in the tight-binding lattice calculation can be implemented by simply removing the impurity site, thus creating a vacancy.%\cite{Mynote5}

\subsubsection*{Self-consistent results}
Above we simply assumed constant order parameters and enforced the correct condensate symmetries. Now we allow the condensate to appropriately respond to the impurity through a proper self-consistent calculation. Since the superconductor symmetry is important for the properties of the bound states, this is the most accurate way to ensure a correct solution.
In these self-consistent calculations we only assume a finite and constant pair potential  $V$ in each pairing channel and calculate the order parameter(s) explicitly everywhere in the lattice. For a  $d$-wave state we use the self-consistent condition  $\Delta_d {(\bf{i},\bf{j})} = -V_d\langle  c_{\bf{i}\downarrow}c_{\bf{j}\uparrow} -   c_{\bf{i}\uparrow}c_{\bf{j}\downarrow} \rangle$, where $\bf{i},\bf{j}$ are nearest neighbour sites. In the self-consistent calculation we start by guessing a value for $\Delta_d$ on each bond, solve Eq.~\eqref{TBH}, calculate a new $\Delta_d$ on each site using the self-consistent condition, and repeat until $\Delta_d$ does not change between two subsequent iterations. For the $d+is$-wave state we also assume a finite $V_s$ in addition to $V_d$ and calculate separately $\Delta_s = -V_s\la c_{\bf{i}\downarrow} c_{\bf{i}\uparrow} \ra$ self-consistently. For the $d+id'$-wave state $V_d'$ is likewise finite, such that $\Delta_{d'} {(\bf{i},\bf{j})} = -V_{d'}\langle  c_{\bf{i}\downarrow}c_{\bf{j}\uparrow} -   
c_{\bf{i}\uparrow}c_{\bf{j}\downarrow} \rangle$, where $\bf{i},\bf{j}$ are next nearest neighbour sites.

We take the initial guess for the subdominant order parameter to be purely imaginary but through the self-consistency loop it is free to acquire a real component as well. Likewise, we emphasize that $\Delta_d$ on $x$- and $y$-directed bonds are treated fully independent and the same 
applies to the $d'$-wave state. Thus, we have not a priori assumed any symmetry for any of the pairing states. In the calculations we use $V_d/t = 1.7$, $V_s/t = 1.7$ for $d+is$-wave state and $V_{d}/t = 1.8$ , $V_{d'}/t = 1.7$ for chiral $d$-wave state, but the results are not sensitive to these particular values. We also set $\mu/t = -1$ for the most general case and to avoid the van Hove singularity at half-filling. We mainly use a $51 \times 51$ lattice, with similar results obtained with a $31 \times 31$ lattice, which guarantees that the result are not sensitive to the lattice size.

\begin{figure}[bht]
\center
\includegraphics[scale=0.65]{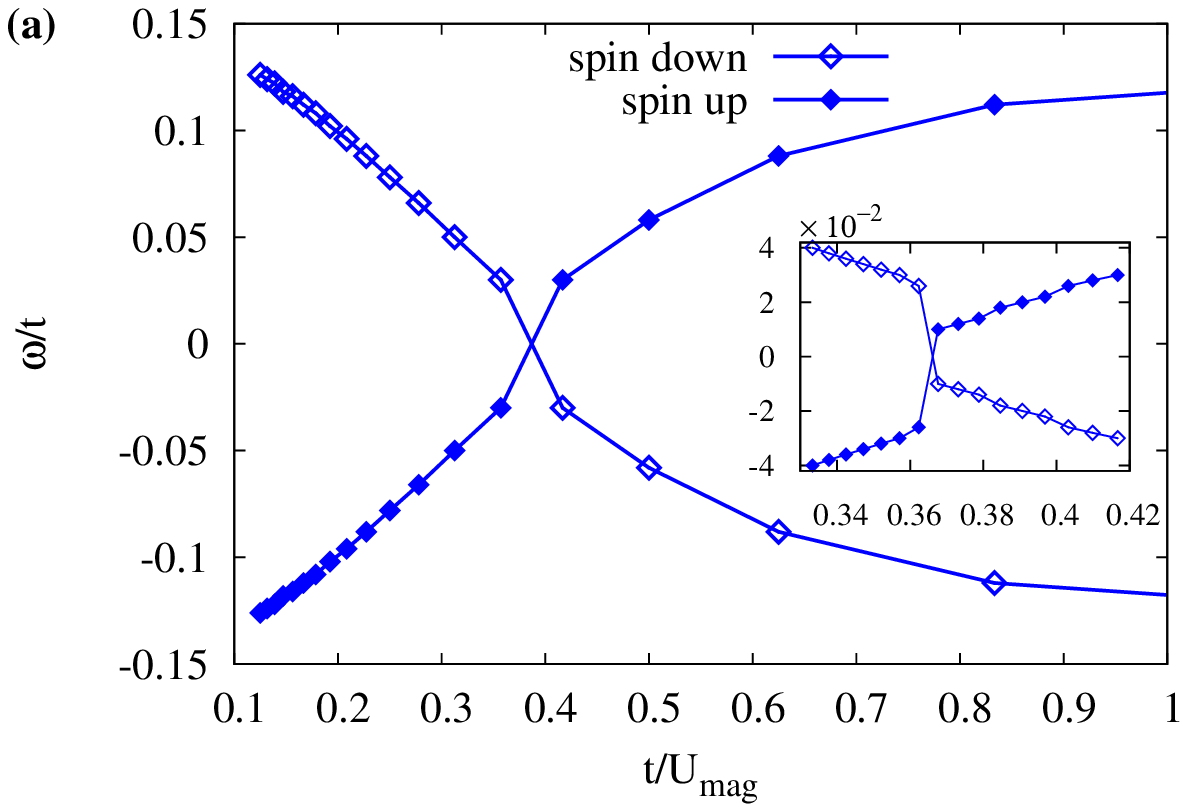}
\includegraphics[scale=0.65]{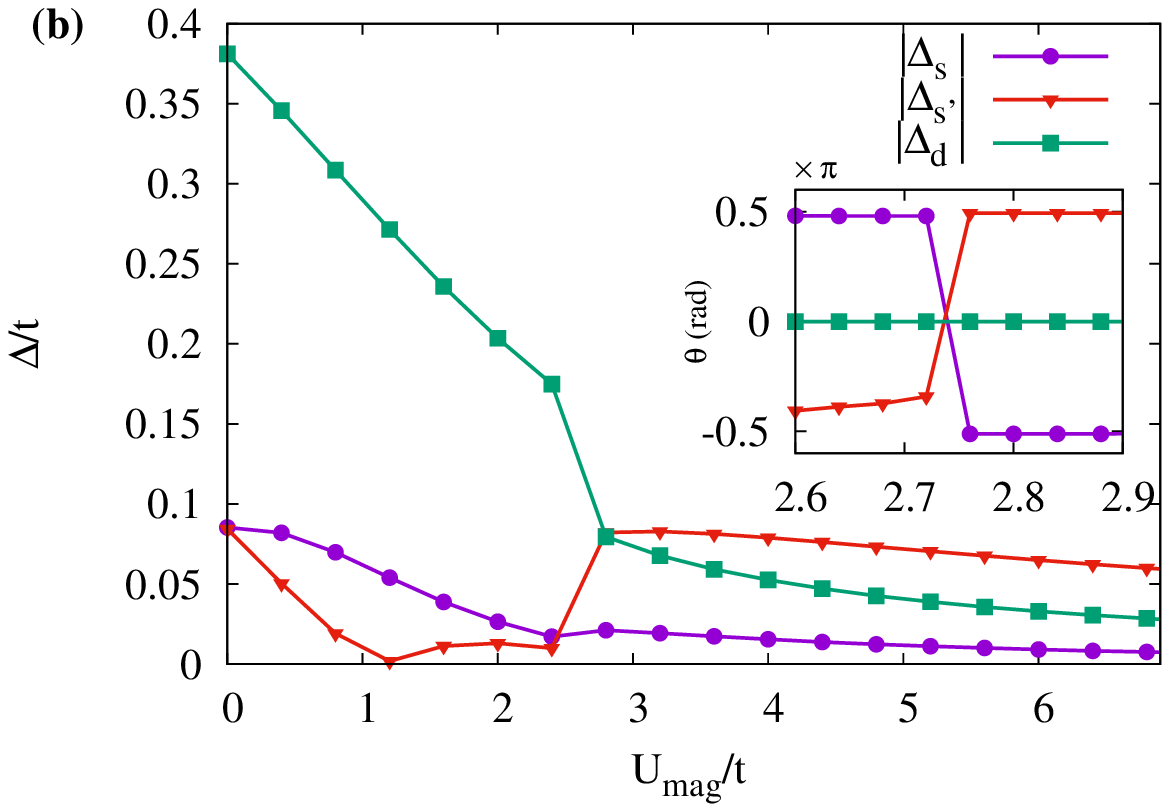}
\caption{The energy spectrum of all intra-gap bound states (a) and the order parameters found self-consistently at the impurity site (b) for a magnetic impurity in a $d_{x^2-y^2}+is$-wave superconductor as a function of impurity strength. Insets shows zoom-ins around the critical point where the zero-energy level crossing, with the inset in (b) showing the phase of the order parameters only.
 }
\label{dis-sc.fig}
\end{figure}

Using the self-consistently calculated order parameters, we extract the LDOS at and close to the impurity site for both $d+is$- and $d+id'$-wave states in the presence of either magnetic or potential impurities.
The self-consistent tight-binding lattice calculations reveal that the results obtained for fixed order parameters are still largely valid.
However, important effects appear around the critical scattering strength in the self-consistent calculations. Starting with the $d+is$-wave state, the self-consistent results show that the intra-gap localized bound states from a magnetic impurity behave largely in a similar way to their non-self-consistent counterpart as seen in Figure~\ref{dis-sc.fig}(a). The main discrepancy is close to the critical scattering $U^c_{mag}$. As seen in the inset in Figure~\ref{dis-sc.fig}(a), the energy of the bound states does not evolve smoothly at the transition point and there is instead a clear kink at $U^c_{mag}$. This is a finger print of a first-order quantum phase transition. A similar effect has been found in a pure $s$-wave superconductor.\cite{SalkolaPRB1997}

The order parameter at the impurity site shed more light on this transition as can be seen in Figure~\ref{dis-sc.fig}(b), where sudden changes near the critical coupling are visible. Self-consistently calculating the order parameters, there is in addition to the on-site $s$-wave and $d_{x^2-y^2}$-wave order parameters, also an extended $s$-wave order parameter. This extended $s$-wave order resides on the nearest neighbour bonds and appears only very close to the impurity. It is thus a direct consequence of the impurity weakening the $d$-wave character in favour of the more disorder-robust $s$-wave symmetry. At the quantum critical point this extended $s$-wave state even becomes the dominant order parameter but notably only at the impurity site, farther away the $d$-wave order parameter is still dominant. 
In addition, both the $s$-wave and extended $s$-wave order parameters develop a $\pi$-phase shift on the impurity site across the critical coupling. This is in line with previous calculations for pure $s$-wave superconductors where the $s$-wave state undergoes a local $\pi$-shift,
\cite{SalkolaPRB1997,Flatte,Kristofer2016PiShift}. However, note that we find that the phase of the dominant $d$-wave state stays constant.

\begin{figure}[htb]
\center
\includegraphics[scale=0.65]{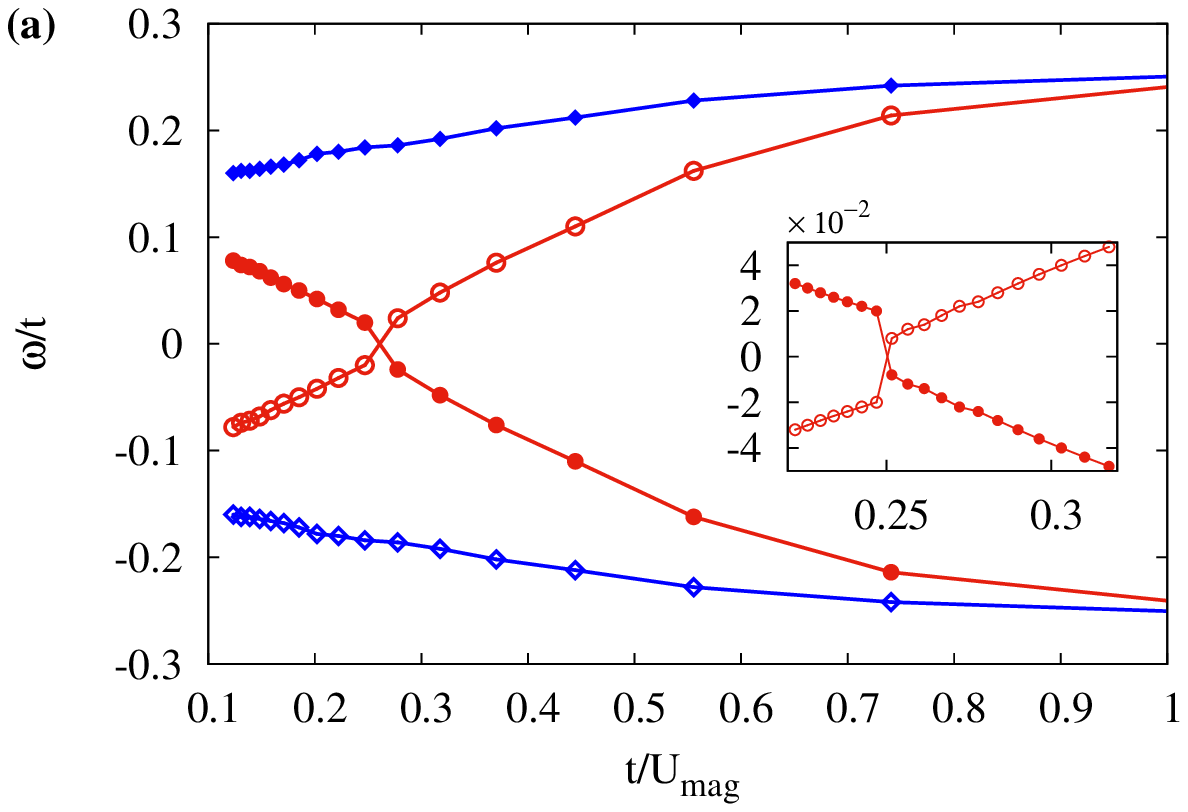}
\includegraphics[scale=0.65]{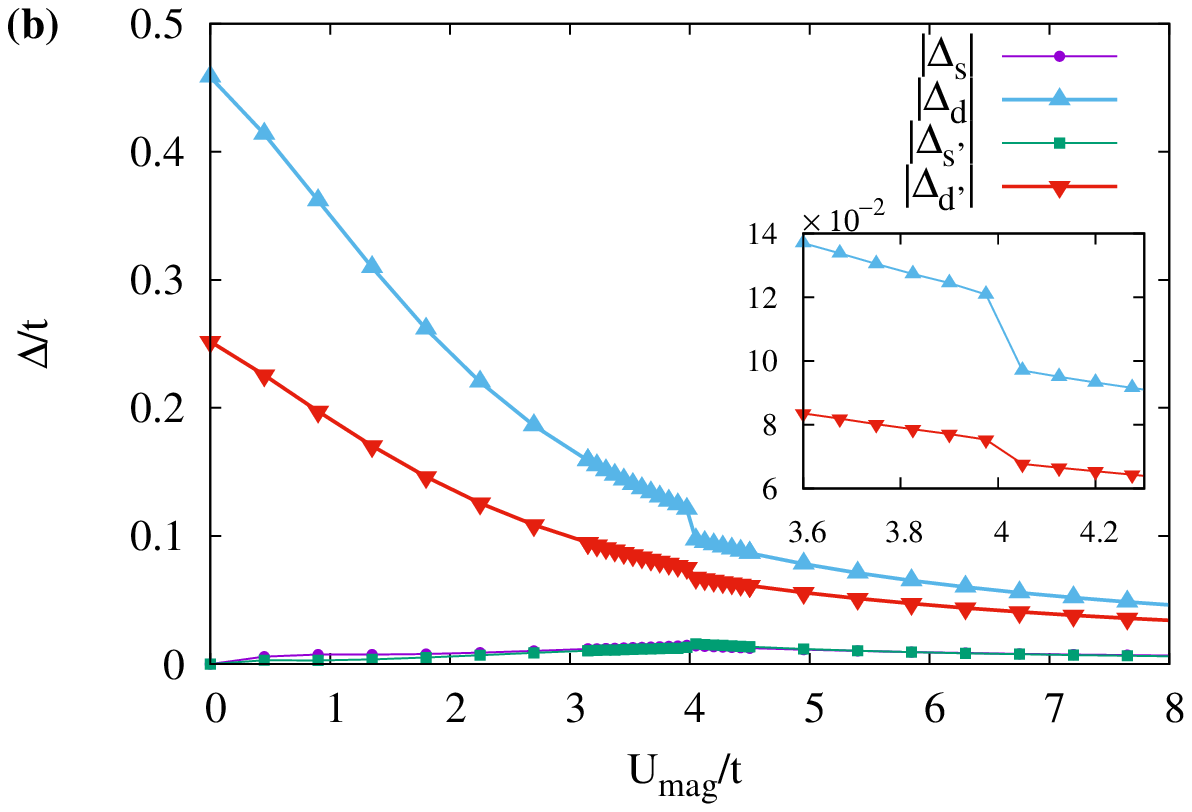}
\caption{The energy spectrum of all intra-gap bound states (a) and the order parameters found self-consistently at the impurity site (b) for a magnetic impurity in a $d_{x^2-y^2}+id_{xy}$-wave superconductor states as function of impurity strength.
Insets shows zoom-ins around the critical point where the energy levels cross zero.
}
\label{LBS-did-sc.fig}
\end{figure}

For the case of a magnetic impurity in a chiral $d$-wave superconductor, there is only a level crossing for one pair of bound states as seen in Figure~\ref{LBS-did-sc.fig}(a), as also found in the non-self-consistent tight-binding and T-matrix calculations. The self-consistent solution, however, shows a kink close to the critical scattering, signaling a first-order phase transition as in the $d+is$ case.
Considering the order parameter, the self-consistent calculation reveals that the dominant $d_{x^2-y^2}$ stays the dominant order parameter 
even beyond the critical scattering and the subdominant state is also always the $d_{xy}$-wave state, as seen in in Figure~\ref{LBS-did-sc.fig}(b). In this case only very weak extended $s$-wave components appear on nearest and next-nearest neighbour bonds, defined here as $\Delta_s$ and $\Delta_{s'}$, respectively. Interestingly, this means that for a chiral $d$-wave superconductor, the impurity does not disturb the dominant $d$-wave orders nearly as much as in the $d+is$-wave case.
Despite the smallness of the $s$-wave components generated in the self-consistent calculations, we find that they are still responsible for lifting of the degeneracy of the impurity bands in the half-filled lattice case $(\mu=0)$.

\section*{Conclusions}
In this work we have investigated impurity-induced bound states in fully gapped $d$-wave superconductors. The main results are summarized in Table \ref{table}. As illustrated by this table, we have shown that an impurity, whether magnetic or potential, induces two pairs of intra-gap bound states in a chiral $d+id'$-wave superconductor, while for a $d+is$-wave superconductor, there is only one (zero) pair of bound states for a magnetic (potential) impurity.
As a result, the number of intra-gap bound states becomes a powerful means for establishing the symmetry of the superconducting state in a fully gapped $d$-wave superconductor, such as that recently established in cuprate nanoislands or at certain cuprate surfaces.\cite{Elhalel,Floriana2013} 
With potential impurities also tunable by localized potential scattering from a tunneling probe,\cite{Sau2013} there even exist possibilities to study the evolution of the bound states of a particular impurity for a range of effective impurity strengths.
\begin{table}[h!]
\begin{center}
\vspace{2ex}
\begin{tabular}{l c c c}
\toprule
\textbf{} & \quad \textbf{$  d+is$} & \qquad \textbf{$    d+id'$} \small{$(\mu=0)$} & \qquad \textbf{$d+id'$} \small{$(\mu\neq 0)$}\\
%\textbf{} &  & $(\mu=0)$ & \textbf{$d+id$}$(\mu\neq 0)$\\
\midrule
$U_{pot}$   	& $\times$ & 2$\times$2 & 2$\times$2 \\
$U_{mag}$   & 2 & 2$\times$2 & 4 \\
$|\tilde{U}^c|$ & $\frac{\pi \sqrt{1+\tilde{\Delta}^2}}{ \ln[16(1+\tilde{\Delta}^2)]}$ & $\infty$ & $\pi \Lambda/|\mu|$\\
$\pi$-shift & $\checkmark$ & $\times$ & $\times$\\
\bottomrule
\end{tabular}
\end{center}
\caption{The  number of bound states and critical scattering strengths for impurities in $d+is$-wave and $d+id'$-wave superconductors. $\tilde{\Delta}$ represents the ratio of dominant over subdominant order parameter, while $2 \times 2$ indicates two states being two-fold degenerate.}
\label{table}
\end{table}

Another important difference between $d+is$-wave and chiral $d$-wave superconductors is the behaviour of the zero-energy level crossings for the impurity-induced states.
For a $d+is$-wave superconductor, increasing the magnetic scattering strength $U_{mag}$ leads to a level crossing of the bound states, which means there always exists a critical coupling $U^{c}_{mag}$, separating two distinct ground states. However, for a chiral $d+id'$-wave superconductor there is no level crossing for a particle-hole symmetric normal band structure (here indicated by $\mu = 0$) at any finite scattering strength, for either potential or magnetic impurities. Only at significant doping away from $\mu = 0$ is there a zero-energy level crossing at an experimentally achievable scattering strength. 
It is also important to notice that the impurity bound states in a chiral $d+id'$-wave superconductor are often twofold degenerate for both potential and magnetic impurities. For a magnetic impurity the degeneracy is lifted by either a finite $\mu$ or by ever-present subdominant extended $s$-wave components, as we find in our self-consistent calculations, nonetheless, there are often two nearly degenerate states. This has important consequences for the generation of zero-energy Majorana modes at the ends of a magnetic impurity wire in chiral $d+id'$-wave superconductors. An even number of (near) zero-energy states in the single impurity limit will result in two putative Majorana end modes for a wire, which then hybridize and split off from zero energy, losing their Majorana character. Thus, in terms of the potential for generating Majorana modes, our results shows that only magnetic impurities in a $d+is$-wave superconductor or in a heavily doped $d+id'$-wave state are promising systems.
Finally, we emphasize that our self-consistent calculations confirm our analytical results where we have assumed constant order parameters uninfluenced by the impurities. In addition, the self-consistent calculations shed more light on the nature of the zero-energy level crossings and show that for both $d+is$ and $d+id'$-wave superconductors, these are first-order quantum phase transitions, with clear discontinuities in both energy levels and order parameters. For the $d+is$-wave state we even find a local $\pi$-shift at the phase transition for all subdominant order parameters, consistent with the behaviour in conventional $s$-wave superconductors.
\cite{SalkolaPRB1997,Flatte} 
For the $d+id'$-wave superconductor we, however, do not find any $\pi$-shifts.

\bibliography{sample}

\begin{thebibliography}{10}
\bibitem{MyNote2}
\bibinfo{author}{{Since $(\Lambda \gg \Delta_2)$, we have $F_3(\omega) \approx
  -\frac{\rho}{2}\ln\left(\frac{(\mu-\Lambda)^2+\Delta^2_2-\omega^2}{(\mu+\Lambda)^2+\Delta^2_2-\omega^2}\right)$.
  Further, if we assume $\mu \ll \Lambda$, then up to first order we find $F_3
  = 2\rho \mu/\Lambda$}} .
\bibitem{Mynote3}
\bibinfo{author}{{The critical coupling for finite but small $\mu$ is
  $\frac{1}{U^c_{mag}} = 2\sqrt{\frac{K^2
  \left(\frac{\tilde{\Delta}^2}{1+\tilde{\Delta}^2}\right)}{\pi^2({1+\tilde{\Delta}^2})}
  + \frac{\mu^2}{\Lambda^2}} $}} .
\bibitem{Mynote4}
\bibinfo{author}{{The main difference between the continuum and discrete
  calculation is that in the former we neglect the radial dependence of the
  order parameter in reciprocal space while in the latter, we keep both the
  radial and angular dependence. Therefore, the discrete summation over the
  full tight-binding model gives us a more relevant solution to compare with
  the finite-size numerical calculations}} .
\end{thebibliography}

\section*{Acknowledgements}
This work was supported by the Swedish Research Council (Vetenskapsr\aa det), the G\"{o}ran Gustafsson Foundation, the Swedish Foundation for Strategic Research (SSF), and the Wallenberg Academy Fellows program through the Knut and Alice Wallenberg Foundation. The computations were performed on resources provided by SNIC at LUNARC.

\section*{Author contributions statement}
M.M. performed all the analytical calculations and together with K.B. the numerical lattice calculations. A.B.S. conceived the idea. M.M. and A.B.S. analysed the results and wrote the manuscript with input from K.B..
\end{document}